\documentstyle[prl,aps,amssymb,amsfonts,twocolumn,epsf]{revtex}
\def\1#1{{\bf #1}}
\def\2#1{{\cal #1}}
\def\4#1{{\tt #1}}
\def\5#1{{\sf #1}}
\def\6#1{{\frak #1}}
\def\7#1{{\Bbb #1}}
\def\8#1{{\rm #1}}
\def\9#1{{\cal #1}}

\newtheorem{The}{Theorem}[section]

\newtheorem{Lemma}[The]{Lemma}

\newtheorem{Cor}[The]{Corollary}
  
\title{\bf Stabilizer codes can be realized as graph codes}
\author{D. Schlingemann \\
 {\small Institut f{\"u}r Mathematische Physik, TU Braunschweig,}\\
  {\small Mendelssohnstr.3, 38106 Braunschweig, Germany.}}
\begin{document}
\draft \maketitle
\narrowtext
\abstract{We establish the connection between a recent new
construction technique for quantum error correcting codes, based
on graphs, and the so-called stabilizer codes: Each stabilizer code
can be realized as a graph code and vice versa.}

\section{Introduction} A well known class of quantum error
correcting codes is the class of stabilizer codes which have
intensively been studied by several authors (e.g.
\cite{CaRainShoSl96a,CaRainShoSl96b}). Some efficient methods for
constructing stabilizer codes  have also been developed (e.g.
\cite{Kn96b,Got96,Got97,BeGra99,BeGra99b,BeGeiGra99}). One problem
with these schemes is, however, that they tend to be rather
subtle, and the verification of their error correcting
capabilities (checking the Knill-Laflamme condition for particular
types of errors \cite{KnLafl95}) often requires rather lengthy
computations.

In a recent paper \cite{SchlWer00} we proposed a new, perhaps
simpler way of constructing quantum error correcting codes, called
graph codes, on which more direct intuitions might be built. Two
basic ingredients are needed: The first is a finite abelian group
whose order is the dimension of the Hilbert space describing a
single elementary quantum system. For the two element group this
is a ``qubit''. The second ingredient of the construction is a
weighted graph with two kinds of vertices, labeling the input and
output systems of the code, respectively.

In the present paper we prove that
each graph code is a stabilizer code and, vice versa, each
stabilizer code has a representation as a graph code.

In this sense the graph code construction is just a new way of
looking at an older construction. However, we believe it will still be
useful, because the intuitions coming with the two ways of representing the codes may
be very different. Part of the appeal of the graph code
construction is that the necessary and sufficient conditions for
error correction are directly "visible" from the structure
of the graph. Useful symmetries for the code can be implemented
by choosing graphs with large symmetry groups compatible with the
error correcting capabilities. These symmetries are not
necessarily the same as the natural symmetries for stabilizer
codes.

The paper is organized as follows: In Section \ref{Stab} we review
the notion of a general stabilizer code. Adjusted to our purpose,
we briefly describe in Section \ref{GR} the concept of a graph
code. In comparison to \cite{SchlWer00} we use here a slightly
more general point of view.
We show in Section \ref{GR-Stab} that every graph code,
is a stabilizer code and how its stabilizer group can be
derived from the graph.
The converse, namely that each stabilizer code has graph code
representation, is proven in Section \ref{Stab-GR}.
Finally we give some concluding remarks in Section \ref{Con}.

\section{Stabilizer codes}
\label{Stab}
We begin by reviewing the notion of
a general stabilizer quantum code
\cite{CaRainShoSl96a,CaRainShoSl96b,Got96,Got97}. For this we need to
introduce some preliminary notions.

Consider a linear space $G$ over a finite field $\7F$.
The dual space is denoted
by $G^*$ and for two vectors $\hat g\in G^*$ and $g\in G$ we write
$\langle\hat g,g\rangle\in \7F$ for the dual pairing.

Concerning the additive structure in $\7F$,
the corresponding dual group $\7F^\wedge$ is isomorphic to
$\7F$ itself. We may choose one group isomorphism $\chi:\7F\to \7F^\wedge$
which is symmetric $\chi(a)(a')=\chi(a')(a)$.
Making use of the multiplicative unit $1$ in $\7F$, we obtain a
character $\varepsilon=\chi(1)\in\7F^\wedge$ and the
prescription
\begin{equation}
G^*\times G\ni(\hat g,g)\to\chi(\hat g|g):=\varepsilon(\langle\hat g,g\rangle)
\end{equation}
yields a non-degenerate bicharacter on $G^*\times G$.

From the physical point of view, the group $G$ represents a
classical configuration space. Having the canonical quantization
procedure in mind, the Hilbert space of the
corresponding quantum system is then given by the complex linear space
$L_2(G)$ of all functions on $G$ with the scalar product
\begin{equation}
\langle \psi,\psi'\rangle=\int \8dg \ \overline{\psi(g)}
\psi'(g)
\end{equation}
where $\int\8d g (\cdot)$ is the normalized Haar measure
on $G$, i.e. the sum over all elements in $G$
divided by the order of $G$.

As far as our subsequent analysis is concerned, the Hilbert space
$L_2(G)$ describes the output system, which is the target system
for encoding the logical bits (or even higher level systems). In
this context a general quantum code can be viewed as an isometric
embedding $\1v\mathpunct:\2K\to L_2(G)$ of a Hilbert space $\2K$
whose dimension is strictly smaller than the order of $G$. The
image of $\1v$ is called the {\em protected subspace}.

The error operations under consideration are generated by
two kinds of natural unitary operations
on $L_2(G)$, namely shift and multiplier, corresponding to bit-flip
and phase errors respectively.
Products and linear combinations of these operators generate the
so called {\em Weyl algebra} over $G$. As basic operations,
Weyl operators $\1w(\hat g,g)$ are products of a shift and a multiplier,
acting on functions $\psi$ in $L_2(G)$ according to the prescription
\begin{equation}
(\1w(\hat g,g)\psi)(g_1)=\chi(\hat g|g_1) \psi(g_1-g)
\end{equation}
where $g$ is an
element of $G$ and $\hat g$ is contained in the dual space $G^*$.

Suppose the error operations, which we wish to correct, are given by
a linear space $\2E$, spanned by a family of Weyl operators
$\{\1w(\hat k_j,k_j)|j\in J\}$. According to the commutation relations for
Weyl operators, this linear space is
invariant under the adjoining Weyl operators. More precisely,
if $E$ is some error operator, belonging to $\2E$, then
$\1w(\hat g,g)E\1w(\hat g,g)^*$ is also contained in $\2E$.
As a consequence, if a quantum code $\1v\mathpunct:\2K\to L_2(G)$ corrects
the errors in $\2E$, then the transformed code
$\1w(\hat g,g)\1v$ has the same capability.
In view of this fact, we call two quantum codes {\em equivalent}
if their corresponding protected subspaces are mapped onto each
other by some Weyl operator.

Considering a linear subspace $S\subset G^*\oplus G$
we obtain an algebra $\6A(G|S)$
which is generated by those Weyl operators
$\1w(\hat g,g)$ for which the pair $(\hat g,g)$
is a member of $S$. Assuming that
$S$ is an isotropic, i.e.
\begin{equation}\label{isotropic}
\langle\hat g_0,g_1 \rangle-\langle \hat g_1,g_0\rangle=0
\end{equation}
holds for all $(\hat g_0,g_0),(\hat g_1,g_1)\in S$,
we are dealing with an abelian algebra
represented on $L_2(G)$. This representation
can be decomposed into irreducible representations (characters).
Thus the Hilbert space $L_2(G)$ is a direct sum
\begin{equation}
L_2(G)=\bigoplus_{\zeta\in\6A(G|S)^\wedge} \2H(\zeta)
\end{equation}
where $\6A(G|S)^\wedge$ is the set of characters on $\6A(G|S)$
\cite{foot2} and $\2H(\zeta)$ is the multiplicity space, carrying
the irreducible representation $\zeta$. By $\1v_{(\zeta,S)}$ we denote
the isometric embedding
of $\2H(\zeta)$ into $L_2(G)$ which
is called the {\it stabilizer code} associated with $(\zeta,S)$
\cite{CaRainShoSl96a,CaRainShoSl96b}. The group generated by Weyl
operators $\1w(\hat g,g)$ with $(\hat g,g)\in S$ is called the
{\em stabilizer group}.

We point out here that the equivalence class of a stabilizer
code associated with
$(\zeta,S)$ only depends on the isotropic subspace $S$.
This can be seen as follows:
Two characters $\zeta_0,\zeta_1$ on $\6A(G|S)$ are related by a
vector $(\hat k,k)\in G^*\oplus G$ according to
\begin{equation}
\zeta_1(\1w(\hat g,g))=\chi(\hat k|g)\chi(\hat g|k)\zeta_0(\1w(\hat
g,g)) \ \ .
\end{equation}
As a consequence, the protected subspace for $(\zeta_0,S)$ is
mapped onto the protected subspace for $(\zeta_1,S)$ by the
Weyl operator $\1w(-\hat k,k)$.

\section{Graph codes}
\label{GR}
For our purpose
we present here a slightly more general concept for graph codes
as it is described in \cite{SchlWer00}. The codes which
we are going to consider here, are determined by the
following objects:
\begin{itemize}
\item
Three linear spaces $H,F$ and $G$ over a finite field $\7F$.
The dimension of $H$ corresponds to the number of input systems,
the dimension of $G$ corresponds to the number of output systems.
\item
As explained later in more detail, the graph
corresponds to a linear
operator $\Gamma\mathpunct:H\oplus F\oplus G\to H^*\oplus F^*\oplus G^*$
which is symmetric, i.e.
$\Gamma^*=\Gamma$ \cite{foot0}.
Let $p_H$, $p_F$, and $p_G$
be the canonical projections onto $H$, $F$  and $G$ respectively. We
require that $\Gamma$ has the
block matrix form
\begin{equation}\label{block}
\Gamma=\left(\begin{array}{ccc}
0&0&B^*\\0&0&C^* \\
B&C&A\end{array}\right)
\end{equation}
with operators
\begin{eqnarray}
&A:=p_G^*\Gamma p_G:G\to G^*
\\
&B:=p_G^*\Gamma p_H:H\to G^*
\\
&C:=p_G^*\Gamma p_F:F\to G^*
\end{eqnarray}
where $B$ is injective.
\end{itemize}

The subsequent analysis, concerning the equivalence of graph and stabilizer codes,
focuses mainly on the symmetric operator $\Gamma$, which
determines the equivalence class of the code completely.
In comparison to \cite{SchlWer00}, we introduce here a formula for the code
which is less explicit, but more suited for the following discussion.

The symmetric operator $\Gamma$ yields an isotropic subspace
\begin{equation}
S_\Gamma=\bigl\{(\Gamma v,v)|v\in H\oplus F\oplus G\bigr\}
\end{equation}
in $H^*\oplus F^*\oplus G^*\oplus H\oplus F\oplus G$
and, by using the notions of the previous paragraph,
we consider a character $\tilde\tau$ of the abelian algebra
$\6A(H\oplus F\oplus G|S_\Gamma)$.
The {\em graph code}, associated with $(\tau,\Gamma)$,
is the linear map $\1v_{(\tau,\Gamma)}:L_2(H)\to L_2(G)$
defined on functions $\psi$ by
\begin{equation}\label{codeformula}
(\1v_{(\tau,\Gamma)} \psi)(h):=\sqrt{|H||F|}
\int \8dh \8df \ \tau(h\oplus f\oplus g) \ \psi(h)
\end{equation}
where $\tau$ is the function on $H\oplus F\oplus G$ given according
to the prescription
\begin{equation}
\tau(v)= \tilde\tau\bigl(\1w(\Gamma v,v)\bigr)
\end{equation}
with $v\in H\oplus F\oplus G$.

Note that, if the graph code $\1v_{(\tau,\Gamma)}$ corrects
$e\geq 0$ errors, then
$\1v_{(\tau,\Gamma)}$ is an isometry.

By a similar argument, as used in the previous section, one
observes that the equivalence class of the graph code
associated with $(\tau,\Gamma)$ only depends on the symmetric
operator $\Gamma$.

For the description of errors, affecting single "bit"
we identify single bits by choosing a basis for each
of the linear spaces $H$, $F$ and $G$.
We take three sets $X$, $J$ and $Y$ with
$|X|=\8{dim}_\7F(H)$, $|J|=\8{dim}_\7F(F)$ and
$|Y|=\8{dim}_\7F(G)$, each labeling a basis:
$e_X=(e_x)_{x\in X}$ is a basis of $H$, $e_J$ a basis of $F$
and $e_Y$ basis of $G$.

These sets correspond to different types of vertices:
The elements of $X$ and $J$ are called "input vertices". They
label the "input systems".  The elements of $Y$ are called "output
vertices", labeling the "output" systems.
As one can see from the expression
(\ref{codeformula}), only the input vertices in $X$
are used for encoding. The inputs in $J$ are used as auxiliary
degrees of freedom for implementing additional constrains for the
protected subspace. According to their role, the elements in $J$
are called "auxiliary vertices".

The errors which affect the output systems,
labeled by elements in some set $E\subset Y$, are linear combinations of
Weyl operators $\1w(\hat g,g)$ where $g$ is contained in the
linear span of $e_E$ and $\hat g$ is a member of the linear span
$e^*_E$ where $e^*_Y:=(e_y^*)_{y\in Y}$ is the dual basis of $e_Y$.

The symmetric operator $\Gamma$ can now be viewed as a weighted
graph on $X\cup J\cup Y$ by declaring
two vertices $z,z'\in X\cup J\cup Y$ to be connected by an edge if
the matrix element
$\Gamma(z,z'):=\langle e_z,\Gamma e_{z'}\rangle\not =0$
is non-vanishing. The value $\Gamma(z,z')\in\7F$ is then the weight
assigned to the corresponding edge.

\subsection{Example}
A simple example for a graph code is given by a quantum code of
length 5, encoding one "bit" and correcting one error.
By choosing $H=\7F$, $F=\{0\}$ (no auxiliary vertices) and $G=\7F^5$
we consider the code which is given
by the symmetric $6\times 6$ matrix
\begin{equation}
\left(\begin{array}{cccccc}0&1&1&1&1&1\\
                       1&0&1&0&0&1\\
                       1&1&0&1&0&0\\
                       1&0&1&0&1&0\\
                       1&0&0&1&0&1\\
                       1&1&0&0&1&0\end{array}\right)
\end{equation}
operating on $\7F^6$. The corresponding graph
is depicted in FIG.1, where the central node, symbolized by  "$\circ$", is
the input vertex.

\begin{figure}[h]
\begin{center}
\epsfysize=2.5cm \epsffile{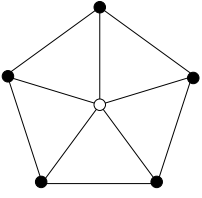}
\end{center}
\caption{Graph code of length five, correcting one error.}
\label{fig1}
\end{figure}

\subsection{Example}
A further example is given by a quantum code of
length 6, encoding one "bit" and correcting one error (FIG.2).
Here we choose $H=\7F$, $F=\7F^3$ and $G=\7F^6$.

There is one relevant input vertex, symbolized by "$\circ$",
three auxiliary vertices, symbolized by "$\otimes$",
and six output vertices "$\bullet$".

\begin{figure}[h]
\begin{center}
\epsfysize=3cm \epsffile{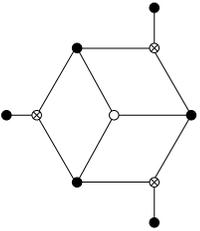}
\end{center}
\caption{Graph code of length six, correcting one error.}
\label{fig2}
\end{figure}

Concerning the example in FIG.2, the
integration over the auxiliary degrees of freedom
(\ref{codeformula}) is nothing else but applying Fourier
transforms to those outputs "$\bullet$" of the graph code in FIG.3
which are not connected with the input "$\circ$".

\begin{figure}[h]
\begin{center}
\epsfysize=2.5cm \epsffile{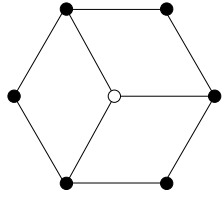}
\end{center}
\caption{Alternative graph code of length six, correcting one error.}
\label{fig3}
\end{figure}

\section{Constructing stabilizer codes from graph codes}
\label{GR-Stab} We are now prepared to show, that each graph code
is indeed (equivalent to) a stabilizer quantum code. It is
convenient to perform the subsequent analysis in  two steps: First
we consider the case, where no auxiliary inputs are needed. Then
we discuss the general case.

\subsection{The case $F=\{0\}$}
We consider now graph codes with no auxiliary inputs, i.e.
$F=\{0\}$. In this case, $\Gamma$ has the form
\begin{equation}\label{block1}
\Gamma=\left(\begin{array}{cc}
0&B^*\\
B&A\end{array}\right)
\end{equation}
and the stabilizer group of the graph code
is given by the following theorem:

\begin{The}\label{theorem1}
A graph code, associated with the symmetric operator
$\Gamma$ (\ref{block1}), is equivalent to
stabilizer codes being associated with the isotropic
subspace
\begin{equation}\label{iso}
S=\bigl\{(A k,k)|k\in\8{ker}(B^*)\bigr\} \ .
\end{equation}
\end{The}
\1{Proof:}
We apply a Weyl operator $\1w(\hat k,k)$,
$(\hat k,k)\in G^*\oplus G$
to the quantum code $\1v_{(\tau,\Gamma)}$
which gives
\begin{eqnarray}
&&\hskip-20pt|H|^{-1/2}\1w(\hat k,k)\1v_{(\tau,\Gamma)} \psi(g)
\nonumber\\
&=&|H|^{-1/2}\chi(\hat k|g) \ \1v_{(\tau,\Gamma)} \psi(g-k)
\nonumber\\
&=&\chi(\hat k-A k|g) \  \tau(k)
\nonumber\\
&\times&
\int_H
\8dh \  \tau(h\oplus g) \ \chi(B^*k|-h) \ \psi(h) \ \ .
\end{eqnarray}
If we only allow for the coding space to pick up a phase factor
which only depends on $(\hat k,k)$, then we have to require
$\hat k =A k$ for all $k$ which satisfy
$B^*k=0$.
Thus the Weyl operator has to be of the form $\1w(A k,k)$
with $k\in\8{ker}(B^*)$.
From this we get
\begin{eqnarray}\label{character}
&&\hskip-20pt
\1w(A k,k)\1v_{(\tau,\Gamma)} \psi
=\tau(k)\ \1v_{(\tau,\Gamma)} \psi \ \ .
\end{eqnarray}
Let $\zeta$ be the character, defined by the prescription
$\1w(A k,k)\mapsto \tau(k)$.
Then we show that the multiplicity space
$\2H(\zeta)$ is precisely the image of quantum code
$\1v_{(\tau,\Gamma)}$. Since the inclusion
$\1v_{(\tau,\Gamma)} L_2(H)\subset \2H(\zeta)$
holds by construction, we only have to check that the dimension
on $\2H(\zeta)$ is $|H|$.
A unitary operator $U$
from $L_2(G/K)$, $K=\8{ker}(B^*)$,
to $\2H(\zeta)$ is given
according to the prescription
\begin{equation}
U\psi(g):=\tau(g) \ \psi([g]_K)
\end{equation}
for $g\in G$, where $[g]_K$ is the equivalence class of $g$ in
$G/K$.
The space $K$ coincides
with the orthogonal complement \cite{ortcomp} of $B(H)$
which has dimension $\8{dim}_\7F(G)-\8{dim}_\7F(B(H))$.
The linear map
$B$ is injective. Thus we find
$\8{dim}_\7F(K)=\8{dim}_\7F(G)-\8{dim}_\7F(H)$ and the quotient space
$G/K$ has dimension  $\8{dim}_\7F(H)$ which implies that the
complex dimension of $\2H(\zeta)$ is $|H|$.
\noindent
$\Box$

\subsection{Example}
Considering the graph code for the graph in FIG.1,
its stabilizer group can directly be derived from
the graph: Assign to each output vertex $y$ (symbolized by "$\bullet$") an element of the
field $g_y\in\7F$ and write them as a column vector. Build a
second column vector $A g$ by assigning to the output vertex $y$
the sum of those $g_{y'}$ for which the vertex $y'$ is connected
with $y$. This yields, for our example, a pair of column vectors
\begin{equation}
(A g,g)=\left(\begin{array}{c}
g_2+g_5\\g_1+g_3\\g_2+g_4\\g_3+g_5\\g_1+g_4\end{array}\right|\left.
\begin{array}{c}g_1\\g_2\\g_3\\g_4\\g_5\end{array}\right) \ \ .
\end{equation}
According to Theorem \ref{theorem1}, the stabilizer group of the
graph code is generated by Weyl operators $\1w(A g,g)$ for which
$g$ fulfills the constraint that
for each input vertex $x$
the sum of all $g_y$ for which $y$ is connected with $x$
is zero, i.e. $\sum_{i=1}^5g_i=0$.
As a consequence, the corresponding isotropic subspace consists of
$|\7F|^4$ elements:
\begin{equation}
\left(\begin{array}{c}
-k_1-k_3-k_4\\k_1+k_3\\k_2+k_4\\-k_1-k_2-k_4\\k_1+k_4\end{array}\right|\left.
\begin{array}{c}k_1\\k_2\\k_3\\k_4\\-k_1-k_2-k_3-k_4\end{array}\right)
\end{equation}
with $k\in\7F^4$. FIG.4 is a graphical representation of an
element of
the stabilizer group corresponding to the choice $k_1=-k_4=-g$ and
$k_2=k_3=0$.
\begin{figure}[h]
\begin{center}
\epsfysize=3cm \epsffile{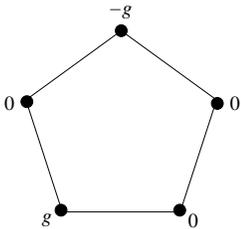}
\end{center}
\caption{Graphical representation of an element of the stabilizer group.}
\label{fig4}
\end{figure}

\subsection{The general case}
Suppose now that the symmetric operator $\Gamma$ is
has the more general form (\ref{block}).
Then we derive from Theorem \ref{theorem1}:

\begin{Cor}\label{cor1}
A graph code, associated with the symmetric operator
$\Gamma$ (\ref{block}), is equivalent to
stabilizer codes being associated with the isotropic
subspace
\begin{equation}\label{iso2}
S=\bigl\{(A k+\hat t,k)|k\in\8{ker}(B^*)\cap\8{ker}(C^*),
\hat t\in\8{ran}(C)\bigr\}
\ .
\end{equation}
\end{Cor}
\1{Proof:}
Let $\Lambda$ be the symmetric operator, mapping $H\oplus G$ into
$H^*\oplus G^*$ which is given by the right hand side of
(\ref{block1}). Consider
the abelian algebra $\6A(H\oplus G|S_\Lambda)$, where
$S_\Lambda$ is the isotropic space $\{(\Lambda w,w)|w\in H\oplus G\}$
and choose a character $\varsigma\mathpunct:\1w(\Lambda
w,w)\mapsto\varsigma(w)$. Then the prescription
\begin{equation}
\tau:\1w(\Gamma(h\oplus f\oplus g),h\oplus f\oplus g)
\mapsto\chi(Cf|g) \ \varsigma(h\oplus g)
\end{equation}
defines a character on $\6A(H\oplus F\oplus G|S_\Gamma)$ and the
corresponding graph code is
\begin{eqnarray}
&&\hskip-20pt(\1v_{(\tau,\Gamma)} \psi)(g)
\\ \nonumber
&=&\sqrt{|H||F|}\int \8dh\8df
\ \chi(f|C^*g)\varsigma(h\oplus g)\psi(h)
\\ \nonumber
&=&\delta_F(C^*g) \ \sqrt{|H|}\int \8dh \ \varsigma(h\oplus g) \ \psi(h)
\end{eqnarray}
where $\delta_F$ is the function
\begin{equation}
\delta_F(f)=\left\{\begin{array}{cc}\sqrt{|F|}&\mbox{ if
$f=0$}\\0&\mbox{ else }\end{array}\right. \ \ .
\end{equation}
Thus
the graph code, associated with $(\tau,\Gamma)$, is
the restriction of the graph code, associated with
$(\varsigma,\Lambda)$ to the kernel of $C^*$:
\begin{eqnarray}
(\1v_{(\tau,\Gamma)} \psi)(g)
=\delta_F(C^*g)(\1v_{(\varsigma,\Lambda)} \psi)(g) \ \ .
\end{eqnarray}
The function with support in $\8{ker}(C^*)$ are precisely those
which are invariant under multiplier operators
$\1w(\hat t,0)$ with $\hat t$ contained in the range of $C$. According to
Theorem \ref{theorem1}, the stabilizer
group of the code $(\tau,\Gamma)$ is generated by Weyl operators
$\1w(A k+ \hat t,k)$ with $k\in\8{ker}(B^*)\cap\8{ker}(C^*)$ and
$\hat t\in \8{ran}(C)$.
$\Box$

\subsection{Example}
Consider the graph code, corresponding to the graph in FIG.2.
Its corresponding symmetric matrix $\Gamma$ has the form
\begin{equation}
\Gamma=\left(\begin{array}{ccc}
0&0&B^*\\0&0&C^* \\
B&C&0\end{array}\right) \ \ .
\end{equation}
According to Corollary \ref{cor1},
its stabilizer group corresponds to the isotropic
space $S=\8{ran}(C)\oplus\8{ker}(B^*)\cap\8{ker}(C^*)$.
It can directly be computed from the graph FIG.2 that
$S$ is the subspace in
$\7F^6\oplus\7F^6$ consisting of elements
\begin{equation}\label{stab10}
\left(\begin{array}{c}
\hat k_1\\ \hat k_2\\ \hat k_3\\ \hat k_1+\hat k_2\\ \hat k_2+\hat k_3
\\ \hat k_3+\hat k_1\end{array}\right|\left.
\begin{array}{c}k_1\\k_2\\-k_1-k_2\\-k_1-k_2\\k_1\\k_2\end{array}\right)
\end{equation}
with $\hat k\in\7F^3$ and $k\in\7F^2$,
where the first three components in (\ref{stab10})
correspond to the output vertices which are not connected with the input.
FIG.5 represents an element of the stabilizer group for the
choice $\hat k_3=\hat g$, $\hat k_1=\hat k_2=0$ and
$k_1=-k_2=g$.

\begin{figure}[h]
\begin{center}
\epsfysize=4cm \epsffile{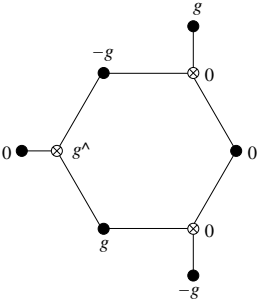}
\end{center}
\caption{Graphical representation of an element of the stabilizer group.}
\label{fig5}
\end{figure}

\section{Constructing graph codes from stabilizer codes}
\label{Stab-GR}
The analysis of the previous section shows that each
graph code
is equivalent to stabilizer quantum code and we can make use of
Theorem \ref{theorem1} and Corollary \ref{cor1}
as well to compute the corresponding
stabilizer group (\ref{iso}) and (\ref{iso2}).
In this section, we show the converse:

\begin{The}\label{Maintheorem}
Each stabilizer code is equivalent to a graph code.
\end{The}

We briefly sketch here the strategy for proving the theorem.

\begin{itemize}
\item
The isotropic spaces which can be obtained from graphs with no
auxiliary inputs (\ref{iso})
are called {\em nondegenerate}. They
are parameterized  by a subspace $K\subset G$ and a linear
map $R\mathpunct:G\to G^*$, which
is symmetric,
such that the isotropic subspace
\begin{equation}\label{nondegenerate}
S=\{(R k,k)|k\in K\}
\end{equation}
is given by the "graph" of $R$, restricted to $K$.
In a first step we consider the nondegenerate case, proving that each
stabilizer code for a nondegenerate isotropic subspace is equivalent
to a graph code (Lemma \ref{lemma0}).
\item
In the second step
we show a general isotropic subspace can be represented as
(\ref{iso2}) (Corollary \ref{cor1}). Making use of this fact,
we prove (Lemma \ref{lemma3}) that a general stabilizer code
has indeed a graph code representation.
\end{itemize}

\subsection{The nondegenerate case}
Consider a nondegenerate
isotropic subspace of $G^*\oplus G$ given by
(\ref{nondegenerate}).
Obviously, the space $G$ is related to the
output system.
Following the proof of Theorem \ref{theorem1}, the
linear space for the input system has dimension
$l=\8{dim}_\7F(G)-\8{dim}_\7F(K)$. A natural choice for the input space
is a linear space $H$ which is isomorphic to $K^\perp$.

\begin{Lemma}\label{lemma0}
Let $V\mathpunct:H\to G^*$ be an injective linear map with
$V(H)=K^\perp$, $K\subset G$, and let $R\mathpunct:G\to G^*$ a symmetric
operator.
A stabilizer code, associated with $S=\{(R k,k)|k\in K\}$,
is equivalent to a graph code, associated
with the symmetric operator
\begin{equation}\label{prorep}
\Gamma=\left(\begin{array}{cc} 0&V^*\\ V&R
\end{array}\right)
\end{equation}
which maps $H\oplus G$ to $H^*\oplus G^*$.
\end{Lemma}
\1{Proof:}
Considering the block matrix form (\ref{block1}) for $\Gamma$, we
identify $A$ with the symmetric operator $R$.
Moreover, we identify $B$ with $V$.
According to Theorem \ref{theorem1} we
conclude that
a graph code associated with $\Gamma$
is equivalent to a stabilizer code
associated with the isotropic group
\begin{equation}
\{(A k,k)|k\in\8{ker}(B^*)\} \ \ .
\end{equation}
Now, we find $\8{ker}(B^*)=V(H)^\perp=(K^\perp)^\perp=K$
and the result follows.
$\Box$

\subsection{Reduction of the general case}
Now, we consider a general isotropic subspace $S\subset G^*\oplus G$.
We introduce the linear subspace $T$
consisting of all elements $t\in G^*$
with $(t,0)\in S$. The subspace $T\oplus\{0\}$ is called the
{\em degenerate part} of $S$. We build the {\em reduced }
isotropic space
$S_\natural:=S/(T\times\{0\})$
as well as the quotient space $G^*_\natural:=G^*/T$.
The equivalence class of a vector $\hat g\in G^*$
in $G^*_\natural$ is denoted by $[\hat g]_\natural$.
Note that the dual space $G_\natural$ of $G^*_\natural$
can be identified with the orthogonal
complement $T^\perp$ of $T$.

In this paragraph, we show that the
reduced isotropic space $S_\natural$ is isotropic and
nondegenerate
in $G^*_\natural\oplus G_\natural$.
For this purpose,
we consider projections
\begin{eqnarray}
&&\hat \pi:S_\natural\ni([\hat g]_\natural,g)
\mapsto[\hat g]_\natural\in G^*_\natural
\nonumber
\\
&&\pi:S_\natural\ni([\hat g]_\natural,g)\to g\in G  \ \ .
\end{eqnarray}

\begin{Lemma}\label{lemma1}
Let $p\mathpunct:G_\natural\to K=\8{ran}(\pi)$ be a projection onto $K$.
Then $S_\natural$ is an isotropic nondegenerate
subspace of $G_\natural^*\oplus G_\natural$ and it is
parameterized by
\begin{equation}\label{param}
S_\natural=\{(R k,k)|k\in K\}
\end{equation}
where
\begin{equation}\label{symm}
R:=\hat\pi\pi^{-1}p+p^*(\hat\pi\pi^{-1})^*(\11-p)
\end{equation}
is a well defined symmetric operator from $G_\natural$ into $G_\natural^*$.
\end{Lemma}
\1{Proof:}
By construction, the projection $\pi$ is injective and on its
range $K$ and $\hat\pi\pi^{-1}$ is a well-defined linear map
from $K$ into $G^*_\natural$.
For $t\in T$ and $(\hat g,g)\in S$ we conclude from the
isotropy of $S$ that $\langle t,g\rangle=0$
which implies $g\in T^\perp=G_\natural$.
Moreover, we obtain for
$([g_0]_\natural,k_0),([g_1]_\natural,k_1)\in S_\natural$ that
\begin{equation}
\langle[g_0]_\natural,k_1\rangle-\langle[g_1]_\natural,k_0\rangle=
\langle g_0,k_1\rangle-\langle g_1,k_0\rangle=0
\end{equation}
is valid for each choice of the representatives $\hat g_0,\hat g_1\in G^*$
since $k_1,k_2$ are contained in $G_\natural$.
Thus $S_\natural$ is isotropic.

Given any vector $([\hat g]_\natural,g)\in S_\natural$,
then $[\hat g]_\natural$ is uniquely determined by
$g\in K$ according to
\begin{equation}
\pi^{-1}g =([\hat g]_\natural,g) \ \ .
\end{equation}
Thus we find $\hat\pi\pi^{-1}g=[\hat g]_\natural$
and the space $S_\natural$ can is given by
\begin{equation}
S_\natural=\{(\hat\pi\pi^{-1}k,k)|k\in K\} \ \ .
\end{equation}

Let $p\mathpunct:G_\natural\to K$ be a projection onto $K$.
Then $p^*\hat\pi\pi^{-1}p$ is a symmetric operator mapping
$G_\natural$ to $G_\natural^*$. Indeed, since $S_\natural$ is
isotropic, we have $\langle k_1,\hat\pi\pi^{-1}k_2\rangle
=\langle k_2,\hat\pi\pi^{-1}k_1\rangle$ for all $k_1,k_2\in K$.
As a consequence the operator
\begin{equation}
R=p^*\hat\pi\pi^{-1}p
+(\11-p^*)\hat\pi\pi^{-1}p+p^*(\hat\pi\pi^{-1})^*(\11-p)
\end{equation}
is symmetric, and we have $R k=\hat\pi\pi^{-1}k$ which finally
implies the identity (\ref{param}).
$\Box$

\begin{Lemma}\label{lemma2}
Let $q\mathpunct:G\to G_\natural$ be a projection onto $G_\natural=T^\perp$,
then the isotropic subspace $S\subset G^*\oplus G$ can be
parameterized as
\begin{equation}\label{generaliso}
S=\{(q^*R k+t,k)|k\in K\subset T^\perp\subset G, t\in T\}
\end{equation}
where $R$ is given by (\ref{symm}).
\end{Lemma}
\1{Proof:}
Let $S'$ be the right hand side of (\ref{generaliso}).
By construction, a vector of the form $(t,0)$, $t\in G$, is
contained in $S'$ iff $t$ is contained in $T$. Now the identity $S'=S$
follows by observing
that $S'/(T\oplus\{0\})$ coincides with the reduced isotropic space
$S_\natural$. Indeed, we have $[q^*Rk+t]_\natural=Rk$
and by Lemma \ref{lemma1} we find $S_\natural=S'/(T\oplus\{0\})$.
$\Box$

In order to formulate the following lemma, we choose two linear
spaces $H$ and $F$, where $H$ is isomorphic
to $K^\perp/T$ and $F$ is isomorphic to $T$.

\begin{Lemma}\label{lemma3}
Let $q$ be a projection onto $G_\natural$,
let $V:H\to G^*$ be an injective linear
map with $V(H)=q^*(K^\perp/T)$ and let $W:F\to G^*$
be an injective linear
map with $W(F)=T$. Adopting the notation above,
then a stabilizer code, associated with $S$, is equivalent to
graph codes, associated with the symmetric operator
\begin{equation}\label{stab-graph}
\Gamma=\left(\begin{array}{ccc}0&0&V^*\\
0&0&W^*\\
V&W&q^*Rq\end{array}\right)
\end{equation}
which maps $H\oplus F\oplus G$ to
$H^*\oplus F^*\oplus G^*$.
\end{Lemma}
\1{Proof:}
Consider a graph code which is given by the symmetric operator
(\ref{stab-graph}). The input systems correspond to $H\cong K^\perp/T$,
the output systems are given by $G$ and, finally, $F\cong T$ is related
to the auxiliary systems. Considering the block matrix form
(\ref{block}) we identify $A=q^*(R p+p^*R^*(\11-p))q$,
$B=V$ and $C=W$ and we conclude from
Corollary \ref{cor1} and Lemma \ref{lemma2}
that each graph code, associated with the
symmetric operator $\Gamma$ (\ref{stab-graph}) is equivalent
to the stabilizer codes, associated with $S$.
$\Box$

\subsection{Example}
Considering the field $\7F_2$ with two elements, then the space of
vectors
\begin{equation}
M:=\{(g,g,g,g)|g\in\7F_2\}\subset \7F^4
\end{equation}
is contained in its orthogonal complement $M\subset M^\perp$ and
therefore $S=M\oplus M$ is an isotropic subspace of
$\7F^4_2\oplus\7F_2^4$. Note that the orthogonal complement of $M$ is given by
\begin{equation}
M^\perp=\{(k_1,k_2,k_3,k_1+k_2+k_3)|k_1,k_2,k_3\in\7F_2\} \ \ .
\end{equation}
Stabilizer codes, associated with a isotropic subspace of this
kind are called self-dual codes.

We apply Lemma \ref{lemma3} to construct an equivalent graph code.
The degenerate part of $S=M\oplus M$ is just given by $M\oplus
\{0\}$ and we obtain for the
reduced isotropic space $S_\natural=\{0\}\oplus M$. Thus,
using the notations of the previous paragraph, we have $T=K=M$
and therefore $R=0$.
The input space can be chosen by $H=\7F_2^2$ which is isomorphic
to the two dimensional space
$K^\perp/T=M^\perp/M$ whose elements are given by
equivalence classes
\begin{equation}
[k]=\{(k_1+g,k_2+g,k_3+g,k_1+k_2+k_3+g)|g\in\7F_2\}
\end{equation}
with $k_1,k_2,k_3\in \7F_2$. There is precisely
one representative in $[k]$ of the form $(0,h_1,h_2,h_1+h_2)$ with
$h_1,h_2\in\7F_2$. Hence an appropriate choice for the injective
map $V:\7F_2^2\to\7F^4_2$ is given by the $2\times 4$ matrix
\begin{equation}
V=\left(\begin{array}{cc}0&0\\1&0\\0&1\\1&1\end{array}\right) \ .
\end{equation}
The auxiliary space $F=\7F_2\cong M$ is simply one dimensional
and we choose $W:\7F_2\to \7F_2^4$ by the $1\times 4$
matrix
\begin{equation}
W=\left(\begin{array}{c}1\\1\\1\\1\end{array}\right) \ .
\end{equation}

\begin{figure}[h]
\begin{center}
\epsfysize=2.5cm \epsffile{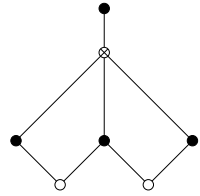}
\end{center}
\caption{Graph constructed from a of a self-dual stabilizer code.}
\label{fig6}
\end{figure}

Applying the identity (\ref{stab-graph}), an equivalent graph code
is associated with the symmetric operator
\begin{equation}
\Gamma=
\left(\begin{array}{ccccccc}0&0&0&0&1&0&1\\
                        0&0&0&0&0&1&1\\
                        0&0&0&1&1&1&1\\
                        0&0&1&0&0&0&0\\
                        1&0&1&0&0&0&0\\
                        0&1&1&0&0&0&0\\
                        1&1&1&0&0&0&0\end{array}\right) \ \ .
\end{equation}
The corresponding graph is depicted in FIG.6, where the inputs are
symbolized by "$\circ$", the auxiliary vertex by "$\otimes$" and the
outputs by "$\bullet$". In particular, the resulting code is an
$[[4,2,2]]$ code, i.e.
it encodes two qubits into four and detects one bit-error.
This can be verified as follows:
\begin{figure}[h]
\begin{center}
\epsfysize=2cm \epsffile{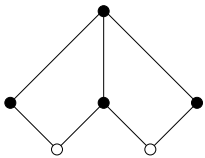}
\end{center}
\caption{Graph of a $[[4,2,2]]$ code.}
\label{fig7}
\end{figure}
Applying a Hadamard transform to the bit, corresponding to the
output vertex on the tip of the graph (FIG.6), one obtains the
graph code, corresponding to FIG.7 which has the same error
correcting capabilities. According to \cite{SchlWer00}
one indeed finds, that the code detects one error.

\section{Conclusion and outlook}
\label{Con}
We have shown that the class of stabilizer codes
coincides with the class of graph codes, where we have assumed
that the degrees of freedom of a quantum register are given by a
linear space over a finite field $\7F$.

As far as the construction of graph codes is concerned, the
degrees of freedom of the input and output register can be described
by any finite abelian group $H$ and $G$ respectively.
The graph code is then determined by a symmetric group homomorphism
$\omega$ mapping $H\times G$ to the dual group $(H\times G)^\wedge$.

The notion of stabilizer codes can also be generalized to
arbitrary finite abelian groups, where a stabilizer code is
determined by an isotropic subgroup $S$ of the direct product of
the dual group $G^\wedge$ with $G$. Here, isotropic means that
for each $(\hat g,g),(\hat h,h)\in S$ the identity
$\hat g(h)=\hat h(g)$ is valid.

By using the
same techniques, as used for the proof of Theorem \ref{theorem1}, one
can show that a graph code, corresponding to any finite abelian
group, is equivalent to a stabilizer code.

Vice versa, applying the methods of the proofs of Theorem \ref{Maintheorem}
by concerning general finite abelian groups, it can be verified that
each stabilizer code
is equivalent to a graph code, provided its isotropic subgroup
$S$ is a {\em retract} in $G^\wedge\times G$, i.e.
there exists a
group homomorphism $p\mathpunct:G^\wedge\times G\to S$ with $p\circ p=p$.

Note that the notion of stabilizer code we have given in Section \ref{Stab}
is related to a linear subspace of a vector space over a finite field.
Since there exists a linear projection onto it,
a linear subspace can also be viewed as a retract,
by only considering the additive structure.

How to construct a logical network, in terms of one and two qubit
operations, which implements a graph code is discussed in a forthcoming paper
\cite{schl01}. The network can directly by derived from the graph.
Since each stabilizer code can be represented by a graph, a
systematic scheme for constructing logical networks for any given stabilizer
code could be developed.

\subsubsection*{{\it Acknowledgment:}}
I am grateful to R.F.Werner for supporting this
investigation with many ideas.
I also acknowledge interesting discussions with A.Klappenecker and
M.Grassl. Funding by the European Union project
EQUIP (contract IST-1999-11053) is
gratefully acknowledged. This research project is also supported by
the Deutsche Forschungsgemeinschaft (DFG-Schwerpunkt
"Quanteninformationsverarbeitung").




\begin{thebibliography}{References}
\bibitem{foot0}
Let $V,W$ be two linear spaces over a field $\7F$.
For an operator $F\mathpunct:V\to W$ the dual map
$F^*\mathpunct:W^*\to V^*$ is determined by
$\langle F^*\hat w,v\rangle:=\langle\hat w,Fv\rangle$ with $\hat
w\in W^*$ and $v\in V$.
\bibitem{foot2}
For an abelian C*-algebra $\6A$, we denote by $\6A^\wedge$ the
collection of all irreducible *-representations (characters)
of $\6A$. This set is a compact Hausdorff space,
called the spectrum of $\6A$,
and $\6A$ can be identified with the continuous functions
on $\6A^\wedge$.
\bibitem{ortcomp}
For a linear spaces $K\subset G$, the orthogonal complement is
the subspace $K^\perp$ consisting of those vectors $\hat g\in G^*$
in the dual space of $G$ for which $\langle\hat g,k\rangle=0$
for each $k\in K$.
\end{thebibliography}
\end{document}